\documentclass[a4paper,12pt]{article} % European paper size
\usepackage{geometry,epsfig,subfigure}
\usepackage{amsmath}
\usepackage{amssymb}
\usepackage{graphics,epsfig,mydef}
\usepackage[square,sort&compress,comma,numbers]{natbib}
\usepackage[usenames]{color}
\usepackage{fancyhdr,url}   % must be loaded for header/footer
\usepackage{setspace}  % must be loaded for \setstretch
\usepackage{multirow}   % must be loaded for \multirow
\usepackage{rotating}    % must be loaded for "sidewaystable" environment
\usepackage{colortbl}     % must be loaded for \columncolor
\usepackage{relsize} % relative sizing of text with smaller/larger
\usepackage{booktabs}
\usepackage{cancel}
\usepackage{enumerate}
\numberwithin{equation}{section}
\usepackage{soul} %for highlights
\usepackage{xcolor} %to cahnge highlight color
\usepackage{pdflscape}
\usepackage{hyperref}
\usepackage[capitalise]{cleveref}
\usepackage{lineno}
\usepackage{graphicx}% Include figure files
\usepackage{etoolbox}
\definecolor{darkblue}{rgb}{0,0,0.8}
\definecolor{darkgreen}{rgb}{0,0.5,0}

\long\def\symbolfootnote[#1]#2{\begingroup \def\thefootnote{\fnsymbol{footnote}}\footnote[#1]{#2} \endgroup} 

\usepackage{amsthm}

\newcommand{\HRule}{\rule{0.9\linewidth}{0.2mm}}

\begin{document}
\renewcommand*{\thepage}{\arabic{page}}

\setstretch{1.3}

\begin{center}
\Large
\textbf{An Energy-Based Interface Detection Method for Phase Change Processes in Nanoconfinements\\}
\normalsize
\vspace{0.2cm}

Mustafa Ozsipahi$^{a}$, Yigit Akkus$^{b}$, Chinh Thanh Nguyen$^{a}$, Ali Beskok$^a\symbolfootnote[1]{e-mail: \texttt{abeskok@smu.edu}}\!$ \\
\smaller
\vspace{0.2cm}
$^a$Southern Methodist University, Dallas, TX 75205, USA \\
%$^b$Department of Mechanical Engineering, Istanbul Technical University, 34437 Gumussuyu, Istanbul, Turkey\\
$^b$ASELSAN Inc., Ankara 06200, Turkey\\
%$^c$University of Illinois at Urbana-Champaign, Urbana, IL 61801, USA\\
\vspace{0.2cm}
\end{center}

\setstretch{1.3} \small

\begin{center} \noindent \HRule \\ \end{center}
\vspace{-0.6cm}
\begin{abstract}

An energy-based liquid-vapor interface detection method is presented using molecular dynamics (MD) simulations of liquid menisci confined between two parallel plates under equilibrium and evaporation/condensation conditions. This method defines the liquid-vapor interface at the location where the kinetic energy of the molecules first exceeds the total potential energy imposed by all neighboring (liquid, vapor, and solid) atoms. This definition naturally adapts to the location of the menisci relative to the walls and can properly model the behavior of the liquid adsorbed layers. Unlike the density cutoff methods frequently used in the literature that suffer from density layering effects, this new method gives smooth and continuous liquid-vapor interfaces in nanoconfinements. Surface tension values calculated from the equilibrium MD simulations match the Young-Laplace equation better when using the radius of curvatures calculated from this method. Overall, this energy-based liquid-vapor interface detection method can be used in studies of nanoscale phase change processes and other relevant applications.

\vspace{0.cm}

\begin{figure}[h]
\includegraphics[width=5.4in]{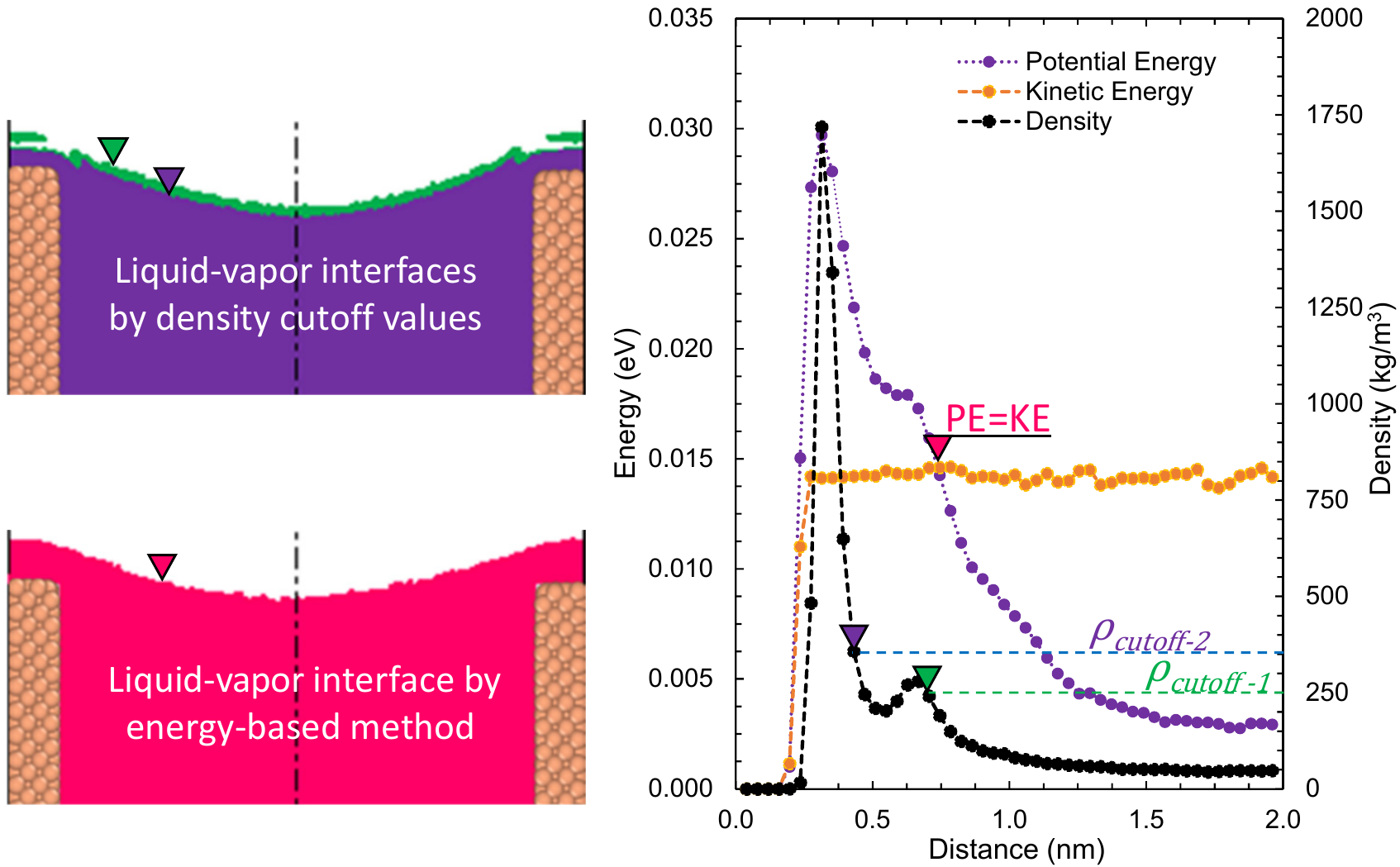}
	\centering	
\end{figure}

%\begin{figure} [h!]
%\includegraphics[width=3.2in]{fig4}
%\centering
%\caption{\label{fig:eqm_rho_vap} Saturated vapor density of argon as a function of temperature. Blue circles are data points and the uncertainty of the data is smaller than the size of the circles. Solid line is the fourth order polynomial fit to the data. Addition of further data points has negligible effect on the polynomial fit.}
%\end{figure}

\vspace{-0.4cm}


\end{abstract}

\vspace{-0.6cm}
\begin{center} \noindent \HRule \\ \end{center}

\pagebreak

Evaporation in nanocapillaries is at the center of several natural and engineered processes such as transpiration of water in plants \cite{wheeler2008}, water desalination \cite{chen2021}, and thermal management of photonic and electronic devices \cite{li2012enhancing}. 
Active mechanisms of evaporation in nanoconfined systems are poorly understood. The dynamics and kinetics of phase transition at the liquid-vapor interfaces at these scales are substantially different from those at the millimeter/micrometer scales due to the increased surface to volume ratio that amplifies the interface and wall force-field effects \cite{li2017, lee2014}. Recent experimental findings exhibit deviations from well-known theoretical expectations such as stretching of the meniscus over the outer surface of the capillary \cite{radha2016}; changes in local evaporation kinetics \cite{li2017,lu2019}; variations in transport resistance of liquids \cite{nazari2019}; evaporation of liquid films spreading outside the capillaries and scale dependent alterations in the evaporation coefficient \cite{li2019}.

Point-wise measurements in nanoconfined liquid-vapor interfaces are challenging. Numerical simulations using molecular dynamics (MD) can be used to investigate transport in these three-phase systems. Evaporation from nanocapillaries is a bubble-free process even at temperatures higher than the boiling temperature, and it occurs under steady-state conditions \cite{nazari2019}. To mimic these conditions and to obtain reliable statistical averaging, MD simulations should be carried out for steady flows. This can be achieved by numerical tricks such as particle removing and injection at the boundaries \cite{rosjorde2000,meland2004noneq} or by creation of a continuously pumped steady flow system \cite{akkus2019molecular,akkus2019atomic}. Regardless of the used approach, determining the exact location of the liquid-vapor interface is challenging. 

In molecular level, the liquid-vapor interface corresponds to a transition region with varying thermophysical properties. The extent of the interfacial region varies based on the thermodynamic state. Even for liquid-vapor mixture of a pure substance well below the critical temperature (\textit{i.e.} low vapor pressure systems), the interface is not perfectly flat, but corrugated by thermal capillary waves \cite{ismail2006}. The corrugated (intrinsic) interface, can be determined by methods focusing on the identification of the molecules forming it \cite{skvor2017}. While early methods using intrinsic sampling or probing \cite{chacon2003,partay2008} were able to identify near-flat surfaces, the subsequent works \cite{sega2013,skvor2017} generalized these methods for arbitrary interface shapes. Employment of these intrinsic analyses are impractical for strongly evaporating interfaces. For such cases, location of the reference surface (Gibbs dividing surface) inside the interfacial region can be determined based on a ``cutoff value" of the sufficiently time-averaged fluid density. Different selection criteria are suggested for the cutoff value, these include the average of the phases' bulk densities  \cite{santiso2013}, Gibbs's equimolar density \cite{rao1979}, half of the bulk liquid density \cite{barisik2013}, arbitrary values between the bulk liquid and vapor densities \cite{akkus2019atomic}, the density value corresponding to the intersection of the extrapolated piece-wise fits applied to the interfacial and bulk vapor region density profiles \cite{akkus2021drifting}. 

In this study, we propose a liquid-vapor interface detection method based on the kinetic and potential energies of fluid atoms. We simply assume that the molecules possessing kinetic energies (KE) greater than the net potential energy (PE) imposed by all surrounding atoms can evaporate. The kinetic energy of an atom equals $\mathrm{KE} =\frac{1}{2} \times m\times v^2$, while its potential energy is highly affected by the neighboring fluid and wall atoms. An energy-based interface definition was \textit{alluded} in \cite{cai2020effects} for investigation of vapor film boiling using \emph{transient} non-equilibrium MD simulations. However, their simulations resulted in a moving flat interface, and the authors did not use this interface definition. 

In order to test the effectiveness of the proposed method, we investigate equilibrium, evaporation and condensation dynamics of a Lennard-Jones fluid (Argon) confined between two parallel Platinum plates with different heights. We specifically utilize the computational setup presented in \cite{akkus2019atomic}, which enables the formation of \textit{steady} liquid-vapor interfaces in the evaporation/condensation sections of the domain using both equilibrium and non-equilibrium MD simulations. The latter approach allows detailed studies of thin-film evaporation and condensation in nanoconfined domains. All molecular dynamics simulations are performed using Large-scale Atomic/Molecular Massively Parallel Simulator (LAMMPS) \cite{plimpton1995}. Computational details are given in the Supporting Information (SI). For all equilibrium MD simulations the thermodynamic state of Argon (Ar) is fixed at $T = 110$\unit{K} and $\rho = 1180\unit{kg/m^3}$ with a quality factor of $x \simeq 0.05$, while all non-equilibrium MD simulations started from this thermodynamic state and ran by equal energy addition and subtraction to/from the wall molecules in the evaporator and condenser sections, respectively.    

\begin{figure}[t]
	\centering	
	\epsfig{file=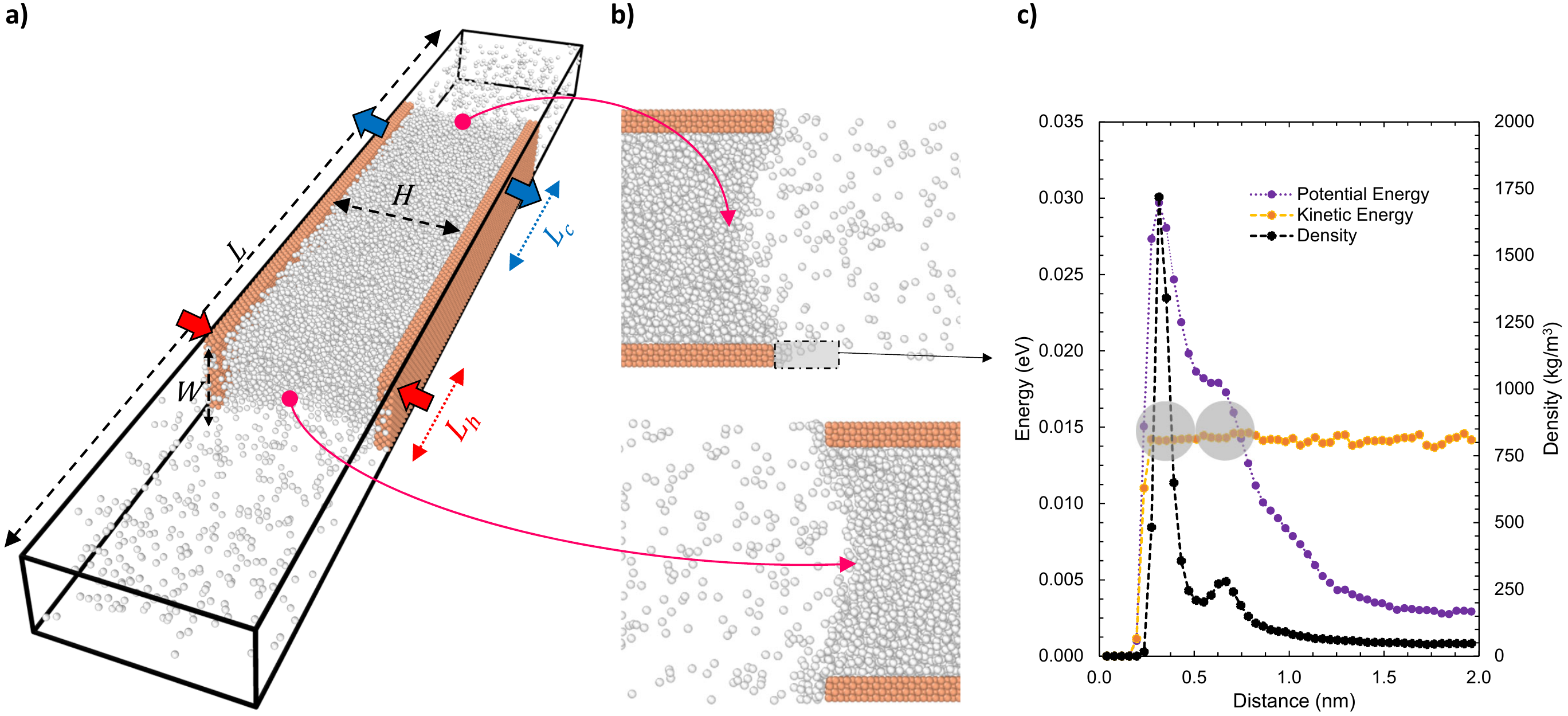,width=6.6in}
	\caption{\label{fig:domain} \textbf{a)} Perspective view of the simulation box of the case M1. Gray spheres represent Ar molecules confined between solid Pt walls (orange spheres). $L_h$ and $L_c$ represent the equal heat extraction/addition regions in the NEMD stage. Wall atoms between the $L_h$ and $L_c$ were frozen in order to eliminate heat conduction between heated/cooled regions of the wall. \textbf{b)} Instantaneous snapshot of two menisci forming at the end of the NVE stage. \textbf{c)} Energy and density distributions of Ar atoms at the near-wall region which is shown as grey box in b). The gray circles in c)  represent atomic diameter of Ar atoms to indicate the scale.}
\end{figure}

Three different geometrically scaled computational domains, from the largest to smallest: \mbox{Model 1} (M1), \mbox{Model 2} (M2), and \mbox{Model 3} (M3), are used to investigate size effects on the new interface detection method. Perspective view of the M1 simulation domain is shown in Fig.~\ref{fig:domain}a. Snapshots of the atomic positions at the liquid-vapor interface during the equilibrium MD simulations are shown in Fig.~\ref{fig:domain}b. The kinetic and potential energy and density distributions of Ar atoms at the side wall of the platinum plate (see the gray box in Fig.~\ref{fig:domain}b) are shown in Fig.~\ref{fig:domain}c, where $x = 0$ corresponds to the lattice position of the last wall atom on the edge of the plate. Time-averaged density and energy values are used to determine the interface thickness. The density profile shows two peaks approximately at $0.3\unit{nm}$ and $0.65\unit{nm}$ from the wall. These two peaks are due to the statistical average of the motion of Ar atoms at respective locations. The first peak is approximately $\sigma_\mathrm{Ar}$ (molecular diameter), and the second peak is $2\sigma_\mathrm{Ar}$ away from the wall. Values of the two density peaks are different than that of liquid Ar confined between the two parallel Pt channels shown in the inset of Fig. 2a, since this zone shows absorbed layer transitioning to the vapor phase at the edge of the plate. For example, the second density peak is at $270\unit{kg/m^3}$, which is about $\mathrm{(1/4)^{th}}$ of the bulk liquid density. Most work in the literature, including our previous studies used an arbitrary density cutoff value to determine the interface. One can choose any cutoff density value between the liquid and vapor Ar densities, and the interface location changes with the chosen density cutoff value. For this specific location, only a density cutoff value smaller than $250\unit{kg/m^{3}}$ will include the second density peak. On the other hand, an energy-based interface is defined at the point where the KE (orange line) fist surpasses the PE (purple line). By this definition, thickness of the interface at this location is predicted slightly larger than two atomic diameters and includes both density peaks from the wall. 

\begin{figure}[t]
	\centering	
	\epsfig{file=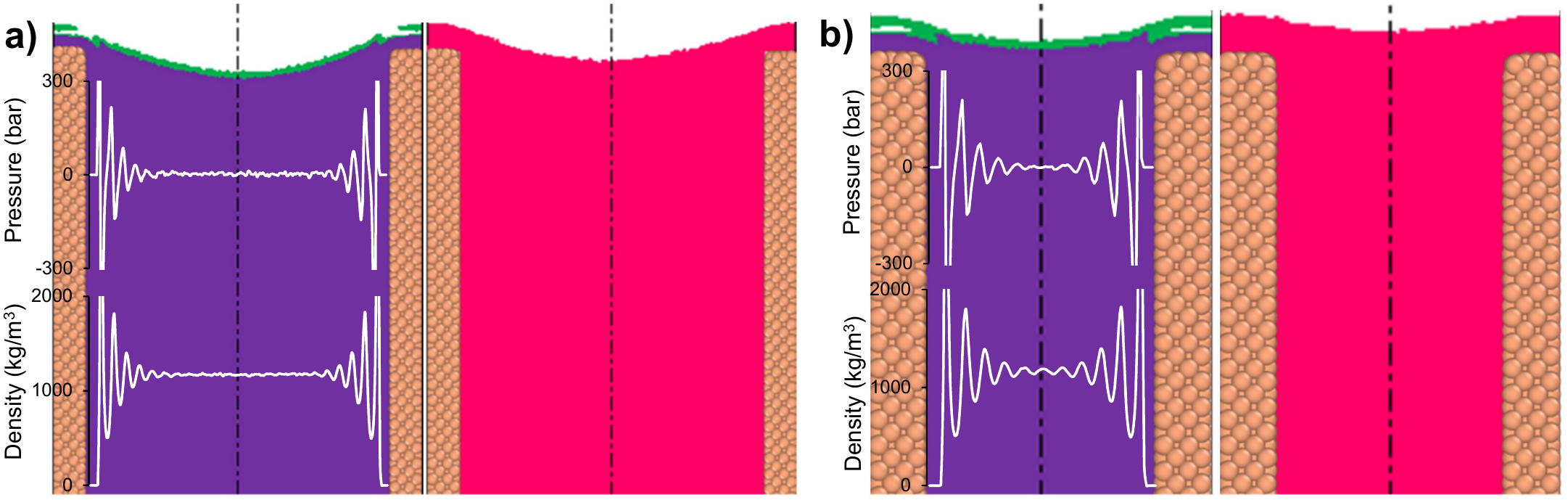,width=6.6in}
	\caption{The liquid/vapor interface profiles in \textbf{a)} M1 and \textbf{b)} M2 during equilibrium simulations. Left figures show the density-based interfaces, where green and purple represent liquid phases obtained using density cutoff values of 250 and 350$\unit{kg/m^{3}}$, respectively. Right figures present the energy-based interface, where red shows the liquid phase. Density and pressure distributions inside the mid-channel are given as insets in left figures. Reducing the channel height increases the surface forces on Ar atoms, resulting in significant fluctuations for both pressure and density profiles.}
	\label{fig:NVE_interface}
\end{figure}

The energy-based interface detection method is implemented by calculating the time-averaged kinetic and potential energies within each bin in the computational domain, and the interface location is identified when the KE first exceeded the local PE. This creates a very sharp interface, since beyond this point KE always exceeds the PE, and the fluid is in vapor phase. Due to the large number of evaporating or condensing atoms, vapor region near the interfaces behaves as dense gas. Density-based interface detection method is employed using the time-averaged density in each bin, and the interface is identified at the location where the density exceeded its ``user decided" cutoff value. Interface predictions under equilibrium MD simulations for M1 and M2 cases using the density cutoff values (250 and $350\unit{kg/m^{3}}$) and the energy-based method are shown in Fig.~\ref{fig:NVE_interface}. Two-identical menisci form on the two edges of the parallel plate system, and only one of them is shown in the figures for brevity. The density-based method using the cutoff value of $250\unit{kg/m^{3}}$ exhibits discontinuity in the liquid-vapor interface region, since this region has two density peaks, as previously shown in Fig.~\ref{fig:domain}c. Furthermore, density layering near the edges of the parallel plate system results in deformation of the meniscus profile around the corner of the walls, and this effect becomes more evident with decreasing the channel height as seen in Fig.~\ref{fig:NVE_interface}b for the M2 case. Insets in Fig.~\ref{fig:NVE_interface} show density and pressure profiles across the liquid section of the channel. The M1 case is large enough to create a constant bulk density and bulk pressure regions, where the M2 case is dominated by the wall force field effects that lead to oscillations. It is important to indicate that pressure in the figures is calculated as the average of the three normal stresses within each bin, and these exhibit anisotropy in the near wall regions, as previously shown in \cite{barisik2011equilibrium}. Overall, the radius of curvature of the meniscus decreases with increasing the density cutoff value. The energy-based interface detection method results in much smoother liquid-vapor interface regions than the density cutoff method and it also predicts a larger meniscus radius.

\begin{figure}[t!]
	\centering	
	\epsfig{file=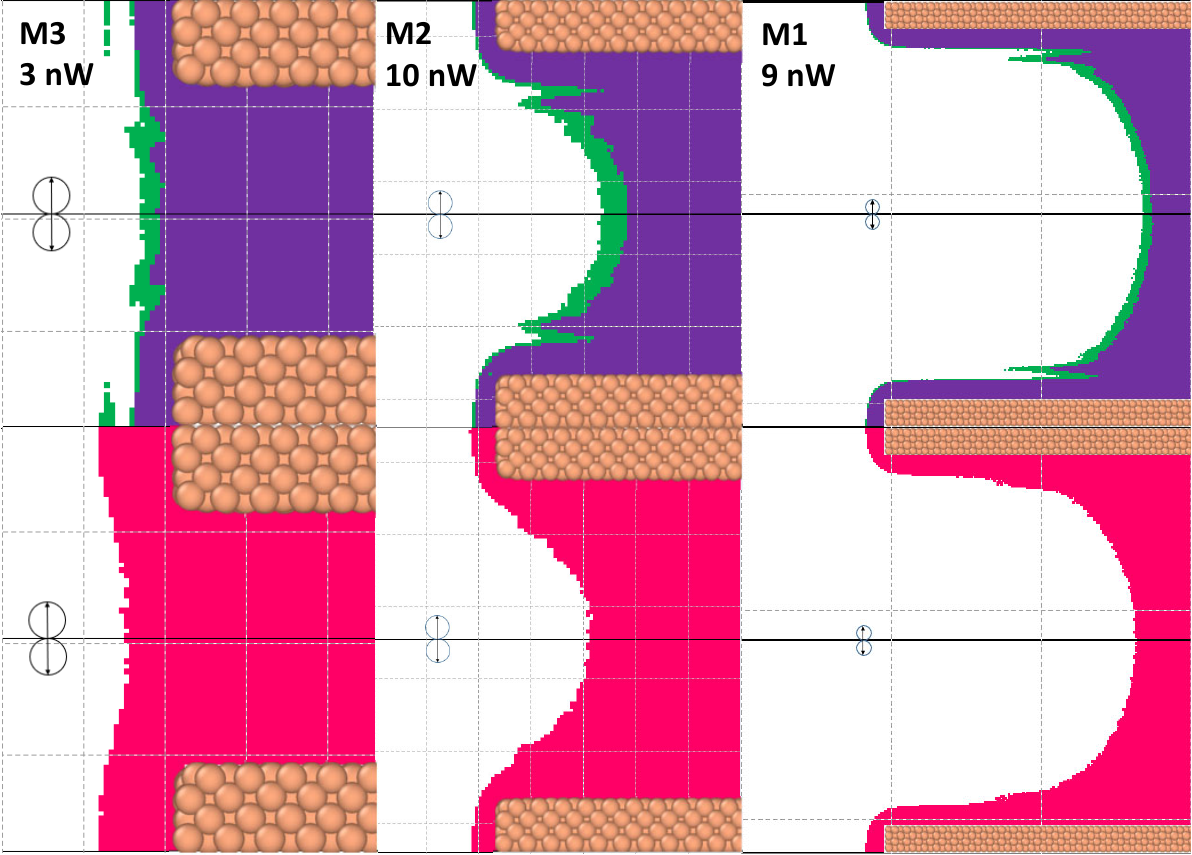,width=5in}
	\caption{Scale effects on the performance of interface detection methods using energy- (bottom row) and density-based approaches (top row, where green and purple represent liquid phases obtained using density cutoff values of 250 and 350 $\unit{kg/m^{3}}$, respectively). Various meniscus profiles are obtained using NEMD simulations employing different H/C values.  14950 Ar atoms are used in M1 to have an elongated interface profile. M3 performed with 1050 Ar atoms. All simulations were performed with $ \epsilon_\mathrm{Ar-Pt}=0.00558\unit{eV}$. A representation of two Ar molecules are shown to indicate the scale. }
	\label{fig:NEMD_E}
\end{figure}

In order to investigate the performance of the energy-based interface detection method under phase change processes, constant heating and cooling are applied in the evaporator and condenser regions of the domain (see Fig.~\ref{fig:domain}a). Moreover, the smallest model (M3) is simulated to observe the scale effects. It should be noted that the simulation setup yields steady-state results, and the interface shapes are averaged using 1,000,000 independent samples. The filling ratio of each simulation domain is carefully selected such that the evaporating meniscus is always placed within the heat addition region of the domain, while the condenser section has mostly flat interface (see Fig.~S1). The menisci in the evaporator regions of the three cases are shown in Fig.~\ref{fig:NEMD_E}. A pinned meniscus forms under 3\unit{nW}heating and cooling (H/C) for the M3 case. The density-based interface detection method is affected by the wall forces and cannot predict a meniscus radius regardless of the density cutoff value. Moreover, adsorbed-layer thickness near the wall edges varies significantly with the used cutoff value for the density-based method. These deficiencies are properly addressed in the energy-based method. The M2 case with 10\unit{nW}H/C exhibits a receded meniscus, which is located nearly at the center of the heated region ($L_h$). The energy-based method presents smooth transition between the meniscus and the adsorbed layer, while the density-based method shows large oscillations for both cutoff values. The M1 case with 9\unit{nW}H/C displays an elongated receded meniscus, where the adsorbed layer formation is obvious in and around the channel. The transition region from the intertwined meniscus to the adsorbed layer is smooth for the energy-based method since it eliminates the fluctuations induced by density layering. 

\begin{figure}[t!]
	\centering	
	\epsfig{file=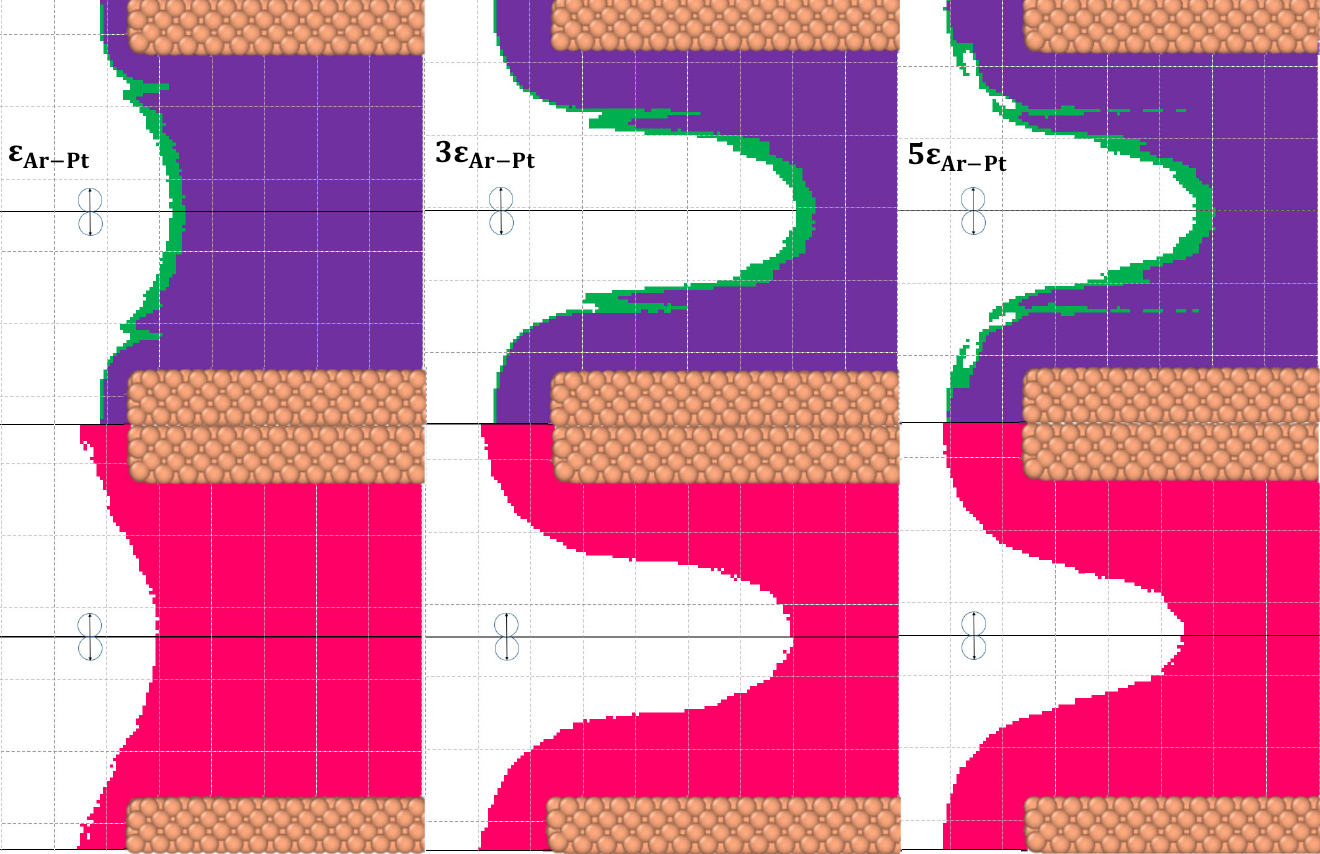,width=5.0in}
	\caption{ The effect of interface detection methods with different wall-fluid interaction potential strength. Tests were performed on M2 using 8\unit{nW}H/C. Top row: Density-based interface definition. Green and purple colors represent the liquid phases obtained using the density cutoff values of 250 and 350 $\unit{kg/m^{3}}$, respectively. Bottom row: Liquid phase obtained using energy-based interface detection method.}
	\label{fig:ewf}
\end{figure}

Next, we investigate the performance of the new interface detection method under increased wall-fluid interaction potentials. Initial studies are performed using Ar-Pt interactions based on the Lorentz-Berthelot mixing rule, resulting in $\epsilon_\mathrm{Ar-Pt}=0.00558\unit{eV}.$ We systematically increased the wall-fluid interaction potential to three and five times larger than this value while applying $8\unit{nW}$ H/C to the M2 case. The comparison of the interface predictions in the evaporator regions under three different interaction strengths are provided in Fig.~\ref{fig:ewf}. Increasing the wall-fluid interaction potential, PE of the Ar atoms increases, while  keeping the same heating value between the three cases resulted in similar KE values (Fig. S2 shows PE, KE, and density variations obtained in the near wall region). For increased interaction potentials, evaporation of Ar from the interface is more difficult and this results in thicker adsorbed layers and smaller meniscus radius. Density-based interfaces at two different cutoff values are oscillatory and discontinuous, and they result in different meniscus curvatures.   

\begin{figure}[t!]
	\centering	
	\epsfig{file=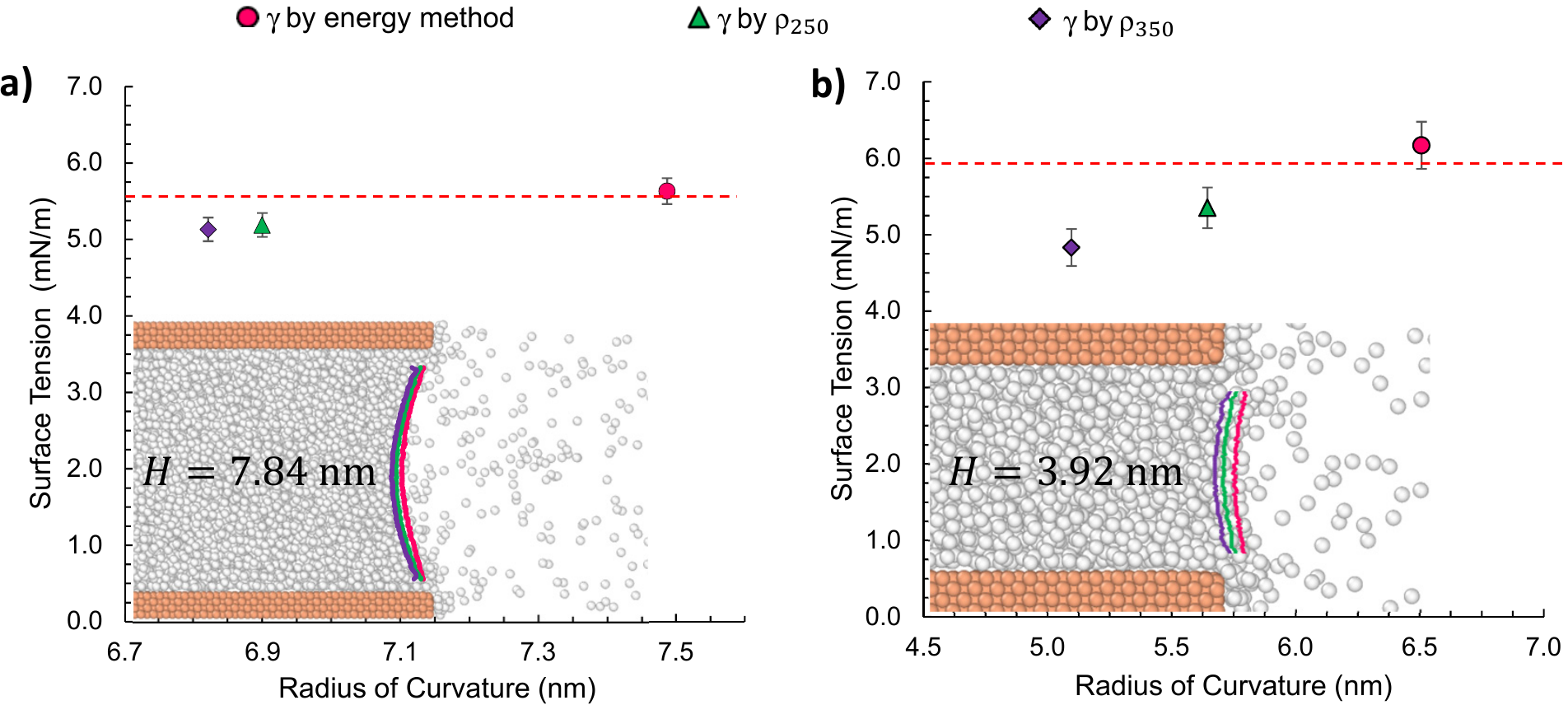,width=6.6in}
	\caption{Young-Laplace (data points) and MD (dashed line) calculations of liquid/vapor surface tension using \textbf{a)} M1 and \textbf{b)} M2. Insets show the resultant interfaces formed at the channel openings, where red line represents the interface obtained by the proposed energy-based method, while green and purple lines represent the ones obtained by density cutoff values of 250 and 350$\unit{kg/m^{3}}$, respectively. Error bars are obtained using various slab thicknesses (group of computational bins) to calculate the pressure difference in Young-Laplace equation.}
	\label{fig:surf}
\end{figure}

The scale, wall-fluid interaction potential, and heating/cooling effects are investigated using the density- and energy-based interface detection methods. We show that the energy-based method yields smoother interface profiles with flatter radius of curvature (ROC). Other physical and thermophysical properties of the interface such as the area, contact angle, temperature, and pressure are used in various models and equations, such as the Hertz-Knudsen, Young’s, Young-Laplace equations. In order to compare the predictions of both interface detection methods, we investigate the consistency of the Young-Laplace equation. For this purpose, we use the Young-Laplace equation that relates the pressure drop at the interface with ROC of the meniscus using surface tension. This requires calculation of the surface tension and the pressure drop between the liquid and vapor sections of the domain from MD simulation data. It is important to indicate that both quantities require the calculation of normal stresses at each computational bin. Pressure is defined as the average normal stress, which becomes isotropic away from the liquid-vapor interface. Surface tension is calculated as an integral effect of anisotropic normal stress components across the same interface (see Section~2 of SI for details). Normal stress calculations in MD utilize Irwing-Kirkwood method, which includes the virial terms that require long-range intermolecular force interactions for accurate calculations \cite{barisik2011equilibrium}. In addition, surface tension in nanoscale can be affected by curvature, capillary wave, and confinement effects \cite{werth2013influence,li2017}. The objective of this study is determination of the advantages of the new energy-based interface detection method, not the exact values of surface tension or pressures. Accurate calculation of these require consideration of several factors described in \cite{mecke1997, kwon2018adhesive, zaleski2020influence, yang2020capillary, kim2021direct}. Here we test the consistency of Young-Laplace equation when using MD calculated surface tension and pressure drop, while the ROC is determined using the density cutoff and energy-based methods. Figure~\ref{fig:surf} presents surface tensions calculated from Young-Laplace equation using ROCs obtained from different interface detection methods (see the insets for interfacial profiles utilized in ROC estimations) during equilibrium MD simulations. Calculation with the ROC obtained from energy-based method yields surface tension values of $\gamma= 5.63$ and $ 6.17\unit{mN/m}$ for M1 (Fig.~\ref{fig:surf}a) and M2 (Fig.~\ref{fig:surf}b) cases, respectively, which are in close agreement with the surface tension values calculated from anisotropic normal stresses (dash lines in Fig.~\ref{fig:surf}) during MD simulations. Therefore, for both M1 and M2 cases, the energy-based interface detection method gives consistent ROCs such that Young-Laplace equation is valid. On the other hand, surface tensions calculated using the ROCs obtained from arbitrarily chosen density cutoff values result in considerable deviations. The extent of deviations  increase with decreasing capillary size.

In summary, a new energy-based interface detection method that identifies the liquid-vapor interface at a location where the averaged kinetic energy first surpasses the local potential energy, is introduced. The performance of this new method is compared with the frequently used density cutoff method using both equilibrium and non-equilibrium MD simulations. Scale, heating and cooling, and wall-fluid interaction potential effects are investigated. Radius of curvatures calculated from both methods are also compared with the predictions from Young-Laplace equation for consistence. In all presented cases the energy-based interface detection method resulted in smooth and superior results than the density cutoff method. Our future studies will use this energy-based interface detection method in investigation of transport in adsorbed layers. This method can also be applied to droplets, liquid bridges, and multi-phase flows and transport in nanochannels.

%%%%%%%%%%%%%%%%%%%%%%%%%%%%%%%%%%%%%%%%%%%%%%%%%%%%%%%%%%%%%%%%%%%%%%

\section*{ASSOCIATED CONTENT}
\subsection*{Supporting Information}
(1) Computational Details; (2) Surface Tension Calculations.
%%%%%%%%%%%%%%%%%%%%%%%%%%%%%%%%%%%%%%%%%%%%%%%%%%%%%%%%%%%%%%%%%%%%%%
\section*{AUTHOR INFORMATION}
\addcontentsline{toc}{section}{Competing interests}
\subsection*{Corresponding Author}
*E-mail: abeskok@smu.edu
\subsection*{ORCID}
Mustafa Ozsipahi: 0000-0003-1378-3991

\noindent Yigit Akkus: 0000-0001-8978-3934

\noindent Chinh Thanh Nguyen: 0000-0003-4868-7414

\noindent Ali Beskok: 0000-0002-8838-5683

\subsection*{Author Contributions}
M.O. performed molecular dynamics simulations. M.O. and Y.A. wrote the manuscript. C.T.N. contributed to the surface-tension calculations. All authors contributed most of the ideas and discussed the results. A.B. reviewed and edited the manuscript.
\subsection*{Notes}
The authors declare no competing financial interests.
%%%%%%%%%%%%%%%%%%%%%%%%%%%%%%%%%%%%%%%%%%%%%%%%%%%%%%%%%%%%%%%%%%%%%%
\section*{ACKNOWLEDGEMENTS}
\addcontentsline{toc}{section}{Acknowledgements}
Mustafa Ozsipahi gratefully acknowledges the financial support by the Turkish Scientific and Technological Research Center under the project 1059B192000153. Computations were carried out using high-performance computing facilities of the Center for Scientific Computation at Southern Methodist University.

\pagebreak

\addcontentsline{toc}{section}{References}
\bibliographystyle{unsrt}
\bibliography{references}

\end{document}